\documentclass[
    amsmath, amssymb,
    aps,
    twocolumn,
    superscriptaddress
]{revtex4-2}

\usepackage{graphicx}
\usepackage{dcolumn}
\usepackage{bm}
\usepackage{booktabs}
\usepackage{tabularx}
\usepackage[colorlinks=true,breaklinks=true,citecolor=magenta]{hyperref}
\usepackage{cleveref}

\usepackage{siunitx}

\DeclareSIUnit \s {\second}
\DeclareSIUnit \ns {\nano\second}
\DeclareSIUnit \mus {\micro\second}
\DeclareSIUnit \ms {\milli\second}
\DeclareSIUnit \MB {\mega\byte}
\DeclareSIUnit \GB {\giga\byte}
\DeclareSIUnit \TB {\tera\byte}
\DeclareSIUnit \PB {\peta\byte}
\DeclareSIUnit \Mbps {\mega\bit/\s}
\DeclareSIUnit \Gbps {\giga\bit/\s}
\DeclareSIUnit \Tbps {\tera\bit/\s}
\DeclareSIUnit \Pbps {\peta\bit/\s}
\DeclareSIUnit \kton {\kilo\tonne} 
\DeclareSIUnit \kt {\kilo\tonne}
\DeclareSIUnit \kty {\kilo\tonne-\year}
\DeclareSIUnit \Mt {\mega\tonne}
\DeclareSIUnit \eV {\electronvolt}
\DeclareSIUnit \keV {\kilo\electronvolt}
\DeclareSIUnit \MeV {\mega\electronvolt}
\DeclareSIUnit \GeV {\giga\electronvolt}
\DeclareSIUnit \TeV {\tera\electronvolt}
\DeclareSIUnit \PeV {\peta\electronvolt}
\DeclareSIUnit \EeV {\exa\electronvolt}
\DeclareSIUnit \sr {sr}
\DeclareSIUnit \m {\meter}
\DeclareSIUnit \cm {\centi\meter}
\DeclareSIUnit \nm {\nano\meter}
\DeclareSIUnit \fm {\femto\meter}
\DeclareSIUnit \in {\inchcommand}
\DeclareSIUnit \km {\kilo\meter}
\DeclareSIUnit \kV {\kilo\volt}
\DeclareSIUnit \kW {\kilo\watt}
\DeclareSIUnit \MW {\mega\watt}
\DeclareSIUnit \MHz {\mega\hertz}
\DeclareSIUnit \mrad {\milli\radian}
\DeclareSIUnit \year {years}
\DeclareSIUnit \POT {POT}
\DeclareSIUnit \sig {$\sigma$}
\DeclareSIUnit\parsec{pc}
\DeclareSIUnit \Gpc {\giga\parsec}
\DeclareSIUnit\lightyear{ly}
\DeclareSIUnit\foot{ft}
\DeclareSIUnit\ft{ft}
\DeclareSIUnit \ppb{ppb}
\DeclareSIUnit \ppt{ppt}
\DeclareSIUnit \samples{S}
\DeclareSIUnit \pe{PE}
\DeclareSIUnit \GeVmwe{GeV/mwe}
\DeclareSIUnit \mwe{mwe}

\newcommand{\enu}{\E_\enu}

\begin{document}
\renewcommand{\MakeTextUppercase}[1]{#1}

\title{Signatures of quasi-Dirac neutrinos in diffuse high-energy astrophysical neutrinos}

\author{Kiara Carloni}
\email{kcarloni@g.harvard.edu}
\affiliation{Department of Physics \& Laboratory for Particle Physics and Cosmology, Harvard University, Cambridge, MA 02138, USA}

\author{Yago~Porto}
\email{yago.porto@ufabc.edu.br}
\affiliation{Centro de Ci\^encias Naturais e Humanas, Universidade Federal do ABC, 09210-170, Santo Andr\'e, SP, Brazil}
\affiliation{{Instituto de F{\'i}sica Gleb Wataghin, Universidade Estadual de Campinas, 13083-859, Campinas, SP, Brazil}}

\author{Carlos~A.~Arg{\"u}elles}
\email{carguelles@fas.harvard.edu}
\affiliation{Department of Physics \& Laboratory for Particle Physics and Cosmology, Harvard University, Cambridge, MA 02138, USA}

\author{P.~S.~Bhupal~Dev}
\email{bdev@wustl.edu}
\affiliation{Department of Physics and McDonnell Center for the Space Sciences, Washington University, St.~Louis, MO 63130, USA}

\author{Sudip~Jana}
\email{hep.sudip@gmail.com}
\affiliation{Harish-Chandra Research Institute, A CI of Homi Bhabha National Institute, Chhatnag Road, Jhunsi, Prayagraj 211 019, India}

\begin{abstract}
    Although the sources of astrophysical neutrinos are still unknown, they are believed to be produced by a population of sources in the distant universe. 
    Measurements of the diffuse, all-sky astrophysical flux can thus be sensitive to flavor and  energy-dependent propagation effects, such as very long baseline oscillations.
    These oscillations are present in certain neutrino mass models, such as when neutrinos are quasi-Dirac.
    Assuming generic models for the source flux, we find that these oscillations can still be resolved even when integrated over wide distributions in source redshift. 
    We use two sets of IceCube all-sky flux measurements, made with muon and all-flavor neutrino samples, to set constraints at the $3\sigma$ level on quasi-Dirac mass-splittings between $(5 \times 10^{-19}, 8 \times 10^{-19})~\textrm{eV}^2$. 
    We also consider systematic uncertainties on the source population and find that our results are robust under alternate spectral hypotheses or physical redshift distributions.
    Our analysis shows that spectral features in the all-sky neutrino measurements provide strong constraints on massive neutrino scenarios and are sensitive to uncharted parameter space.
\end{abstract}
\maketitle

\section{Introduction.}
The fundamental origin of neutrino oscillations and the precise nature of neutrino mass remain unknown. 
Currently, the best bet we have to resolve the Dirac versus Majorana nature of neutrinos is by observing the neutrinoless double beta decay ($0\nu\beta\beta$) process~\cite{Agostini:2022zub}. 
However, there is no guarantee that even next-generation ton-scale $0\nu\beta\beta$ experiments will yield a positive signal if neutrino masses follow the normal ordering, a scenario that is mildly favored by current global oscillation fits \cite{Esteban:2024eli}.
On the other hand, traditional oscillation-based neutrino experiments are only sensitive to the squared mass differences between generations, and do not alter chirality; hence, they cannot resolve the distinction between Dirac and Majorana neutrinos. 
As we will show below, however, measurements of astrophysical neutrino spectra are sensitive to the oscillations of high-energy neutrinos produced by extremely small squared-mass differences, and can thus test a whole class of neutrino mass models with tiny lepton number violation. In fact, this could be the first testable prediction of a class of string theory landscape constructions. 

\begin{figure}[b!]
    \centering \includegraphics[width=\linewidth]{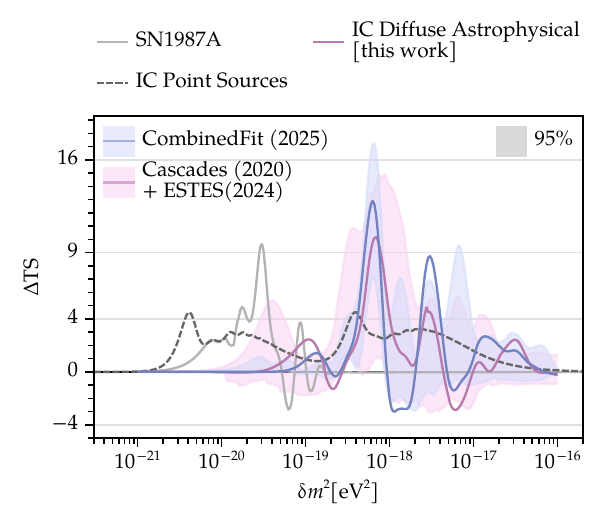}
    \caption{
        \textbf{\emph{Constraints on the Quasi-Dirac parameter space.}} 
        The test statistic difference with respect to the broken power-law null hypothesis, as a function of the QD mass-squared difference $\delta m^2$, is plotted for analyses based on IceCube's 2025 \texttt{CombinedFit}~\cite{IceCube:2025ewu} (blue) and on a combination of \texttt{Cascades} (2020)~\cite{IceCube:2020acn} and \texttt{ESTES} (2024)~\cite{IceCube:2024fxo} results (purple). 
        Brazil bands indicate the regions contained by 95\% of realizations drawn from the best-fit BPL spectrum in~\cite{IceCube:2025ewu}. 
        Previous constraints from Ref.~\cite{Martinez-Soler:2021unz} and sensitivities from Ref.~\cite{Carloni:2022cqz} are shown in grey.
    }
    \label{fig:main}
\end{figure}

If right-handed neutrino fields $\nu_R$ exist, neutrinos may acquire their mass in the same way the charged fermions do, e.g., via a Dirac mass term originating from a Yukawa coupling to the Higgs field.
However, since these new fields would be singlets under the Standard Model (SM) gauge group, no symmetry would then forbid a Majorana mass term for the right-handed fields. 
Since, in a bottom-up phenomenological approach, the Majorana mass $m_R$ is completely unconstrained, it is theoretically allowed to be much smaller than the original Dirac mass term~\cite{Wolfenstein:1981kw,Petcov:1982ya, Valle:1983dk,Doi:1983wu,Kobayashi:2000md}.
This scenario produces what are called quasi-Dirac (QD) neutrinos, and is the focus of this work.

The only experimental way to directly probe QD models is by searching for oscillations in the spectra of neutrinos from astrophysical sources, such as the Sun~\cite{Giunti:1992hk, Cirelli:2004cz, deGouvea:2009fp,Anamiati:2017rxw, deGouvea:2021ymm, Ansarifard:2022kvy, Franklin:2023diy}, supernovae~\cite{DeGouvea:2020ang, Martinez-Soler:2021unz}, high-energy cosmic sources~\cite{2000ApJS,2002ApJS, Beacom:2003eu, Keranen:2003xd, Esmaili:2009fk, Esmaili:2012ac, Joshipura:2013yba, Shoemaker:2015qul, Brdar:2018tce, Carloni:2022cqz, Fong:2024mqz, Dev:2024yrg}, or relic neutrinos~\cite{Perez-Gonzalez:2023llw}. 
These oscillations are driven by hyperfine mass-squared differences, $\delta m^2_k \propto m_R$.
Stringent upper limits on $\delta m^2_{1,2}\lesssim 10^{-12}\si\eV^2$ have been derived using solar neutrinos~\cite{deGouvea:2009fp, Ansarifard:2022kvy}. 
These solar limits supersede a previously derived constraint from Big Bang Nucleosynthesis (BBN), $\delta m^2_k\lesssim 10^{-8}\si\eV^2$~\cite{Barbieri:1989ti, Enqvist:1990ek}.
Both these limits are derived assuming maximal active-sterile neutrino mixing in the QD scenario. 
If the mixing is non-maximal, the solar $\delta m^2$ limits can be much weaker~\cite{Chen:2022zts, Ansarifard:2022kvy}. 
Moreover, the solar neutrino data is not sensitive to $\delta m^2_3$ due to the small electron component in $\nu_3$, and the limits from atmospheric, accelerator and reactor neutrino data are very weak, $\delta m^2_3\lesssim 10^{-5}\si\eV^2$~\cite{Anamiati:2017rxw}, due to the much shorter baselines. 

The identification of a few point sources of astrophysical neutrinos~\cite{IceCube:2022der} provides sensitivity to QD neutrinos with $\delta m^2_k\in [10^{-21},10^{-16}]\si\eV^2$~\cite{Carloni:2022cqz}; see also Refs.~\cite{Rink:2022nvw, Dixit:2024ldv} for related analyses.
However, the current prospective for exploring the QD parameter space in this mass range using point sources is limited by the small number of identified sources, poor resolution on the muon neutrino energy, and lack of flavor information.
These problems do not allow high-significance statements about QD neutrinos using point sources at present.

In this work, for the first time, we use the diffuse astrophysical neutrino flux measured by IceCube in multiple channels and flavors to explore QD models with mass-squared differences $\delta m^2_k \in [10^{-21}, 10^{-16}]\si\eV^2$. 
This strategy has two advantages over point-source based analyses: firstly, measurements in multiple channels with different flavor combinations are available, and secondly, the astrophysical neutrino sample size is much larger.
Furthermore, as we will show, for physically motivated choices of the spatial distribution of astrophysical sources, the QD oscillation probability is not fully smeared out.
It is therefore possible to search for oscillation-induced disappearance dips in the astrophysical neutrino spectrum.

Our main results are shown in~\Cref{fig:main}. 
Using two different sets of IceCube measurements, we find that the absence of disappearance signatures in the spectrum around $10-100~\textrm{TeV}$ allows us to place the first significant constraints on the QD mass-squared difference $\delta m^2$ from $(5-8) \times 10^{-18}~\textrm{eV}^2$.

\section{Theory of quasi-Dirac neutrinos.}
It is unknown whether the neutrino mass term is Majorana, Dirac, or a mixture of both~\cite{Wolfenstein:1981kw,Petcov:1982ya}.
In particular, if the Majorana mass term is much smaller than the Dirac mass, neutrinos are quasi-Dirac~\cite{Valle:1983dk,Doi:1983wu,Kobayashi:2000md}, which are fundamentally Majorana fermions but behave like Dirac in most experimental settings. 
In the QD scenario, lepton number is only slightly violated, and lepton number-violating observables such as $0\nu\beta\beta$ are strongly suppressed.
Ordinary neutrinos have three generations of mass states, $\nu_i$, whose decomposition into flavor states, $\nu_\alpha$, is described by the PMNS matrix.
If neutrinos are QD, each generation has a further hyper-fine mass splitting, $\delta m^2_k$, which produces two mass states per generation, $\nu_k^{\pm}$, which are orthogonal equal combinations of active (LH) and sterile (RH) states. 
In the same way that ordinary neutrinos undergo flavor oscillations after traveling distances $L \sim E/\Delta m^2$, QD neutrinos undergo active-sterile oscillations over much longer distances $L \sim E/\delta m^2$, where $E$ is the neutrino energy.
These ultra-long baseline oscillations are the only distinguishing signature of quasi-Dirac neutrinos. 

The theoretical and model-building aspects of QD neutrinos have been extensively discussed in the literature; see e.g., Refs.~\cite{Chang:1999pb, Nir:2000xn,Joshipura:2000ts,Lindner:2001hr, Balaji:2001fi,Stephenson:2004wv,McDonald:2004qx, deGouvea:2009fp, Ahn:2016hhq, Joshipura:2013yba, Babu:2022ikf, Carloni:2022cqz}. 
It is interesting to note that certain string theory landscape constructions, such as Swampland, predict that neutrinos are Dirac-like particles~\cite{Ooguri:2016pdq,Ibanez:2017kvh,Gonzalo:2021zsp,Casas:2024clw}.
Additionally, we expect that global symmetries such as lepton number are ultimately broken by quantum gravity, turning Dirac neutrinos into quasi-Dirac.
In fact, any model where neutrinos start as Dirac particles with conserved lepton number could receive non-renormalizable quantum gravity corrections which generate small $\delta m^2$  via higher-dimensional lepton-number-violating operators suppressed by the Planck scale, thus making neutrinos naturally quasi-Dirac. 
QD neutrinos could also help explain some features of astrophysical and cosmological observations.
For example, small $\delta m^2$ values could also be linked to the observed baryon asymmetry of the Universe~\cite{Ahn:2016hhq, Fong:2020smz}.
QD neutrinos have also been suggested as solutions to the excess in the diffuse radio background~\cite{Chianese:2018luo, Dev:2023wel}. 

%
\begin{figure*}[ht!]
    \centering \includegraphics[width=\textwidth]{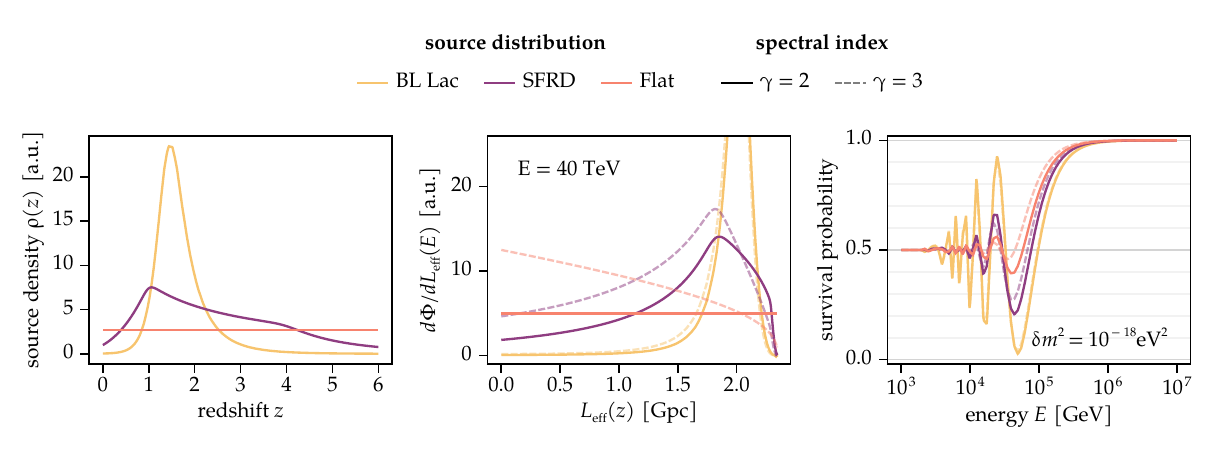}
    \caption{
        \textbf{\emph{Source distribution and integrated oscillation probability.}} 
        \emph{Left:} Three choices for the source distribution in redshift: a flat distribution, the SFRD of Ref.~\cite{Elias-Chavez:2018dru}, and the distribution of BL~Lacs from Ref.~\cite{Groth:2025aan}.
        \emph{Middle:} Distribution of neutrinos with energy $E = \SI{40}{TeV}$ over the effective distance ($L_\textrm{eff}$), assuming a given redshift distribution $\rho(z)$ and power-law emission spectrum with index $\gamma$.
        \emph{Right:} QD survival probability as a function of detected neutrino energy, integrated over the population of sources at all redshifts.
    }
    \label{fig:oscs}
\end{figure*}
\section{Quasi-Dirac oscillations on astrophysical scales.}
A quasi-Dirac neutrino mass model modifies standard flavor oscillations by introducing additional frequencies beyond those associated with the ``solar" ($\Delta m^2_{21} \approx 7.5 \times 10^{-5}\text{eV}^2$) and ``atmospheric" ($\Delta m^2_{31} \approx 2.5 \times 10^{-3}\text{eV}^2$) mass-squared differences. 
These new oscillation components convert mostly active neutrino mass eigenstates, $\nu_k$, into their sterile counterparts with frequencies proportional to $\delta m^2_k$.
For any combination of neutrino energy, $E$, and propagation distance, $L$, suitable for probing allowed QD mass squared differences, $\delta m^2$, the standard oscillation terms containing $\Delta m^2$ will average out due to limited experimental resolution. 
In this regime, the oscillation probability is given by~\cite{Carloni:2022cqz}
\begin{equation} \label{eq:prob}
P_{\alpha \beta}=\frac{1}{2} \sum_{k=1}^3\left|U^*_{\beta k} U_{\alpha k}\right|^2\left[1+\cos \left(\frac{\delta m_k^2 L_{\mathrm{eff}}}{2 E}\right)\right].
\end{equation}
Here, $E$ is the neutrino energy as measured on Earth, and
\begin{equation} \label{eq:Leff}
L_{\mathrm{eff}}= \int \frac{d z}{H(z) (1+z)^2},
\end{equation}
is the effective distance traversed, which accounts for neutrino propagation in an expanding Universe and depends on the Hubble expansion rate, $H(z)$ (see~\cite{coh_footnote} for a discussion about decoherence effects).

The diffuse neutrino flux is produced by a population of sources distributed across cosmological time and space. 
Without specifying the exact nature of these sources, we assume that they are distributed in redshift according to some function $\rho(z)$, e.g., proportional to the star formation rate density (SFRD). 
In this scenario, the number of neutrino sources within a redshift interval $[z, z+dz]$ is given by \cite{Capel:2020txc}
\begin{align}
dN(z) = {\rho}(z) \times 4\pi D^2 \frac{dD}{dz} dz,
\end{align}
where $D(z) = \int dz/H(z)$ is the comoving distance.

In this work, we assume that sources at all redshifts emit neutrinos with the same spectrum and in the same ratio of flavors.
In this case, the flux of neutrinos of flavor $\alpha$ at Earth from a source at redshift $z$ is given by 
\begin{align}
\Phi_\alpha(E,z) = f_\alpha\;  \phi_0\left( E(1+z) \right) \times \frac{(1+z)^2}{4\pi D_L^2}.
\end{align}
In this expression, $f_\alpha$ is the fraction of neutrinos produced in flavor $\alpha$, and $\phi_0$ is the common functional form for the emitted neutrino spectra. 

Including QD oscillations, the contribution to the total diffuse flux from a given redshift interval is
\begin{equation}
d\Phi_\beta(z,E)= \sum_\alpha P_{\alpha \beta}(E,z) \Phi_\alpha(E,z)\; dN(z).
\end{equation}
Thus, the total flux at Earth as a function of energy is therefore given by the integral over redshift:
\begin{multline} 
\Phi_\beta(E) = \int dz \sum_\alpha P_{\alpha \beta}(E,z) \times f_\alpha \phi_0(E(1+z)) \times \frac{\rho(z)}{H(z)}.
\label{eq:flux_avg}
\end{multline}

In~\Cref{fig:oscs}, we show a range of possibilities for the source redshift distribution $\rho(z)$: a flat distribution in $z$, the star formation rate density function from Ref.~\cite{Elias-Chavez:2018dru}, which grows for small $z$ as $(1+z)^3$, and the distribution of BL~Lacs from Ref.~\cite{Groth:2025aan}, which grows for small $z$ as $(1+z)^5$.
These distributions all correspond to mean effective distances $L_\textrm{eff}$ on the order of Gpc, and thus produce first oscillation minima at tens of TeV for a mass-squared difference $\delta m^2 = 10^{-18}\;\si{\eV\squared}$.
The SFRD and BL~Lac distributions, which grow steeply with $z$, produce a much more pronounced disappearance effect than the flat distribution.
The oscillation curve also changes with different choices of the source emission spectrum $\phi$.
A softer flux, with a power-law spectral index $\gamma = 3$ instead of 2, produces a smaller disappearance signature, since more of the neutrinos at a given energy $E$ will be coming from nearby sources.


\section{Analysis.}
Quasi-Dirac oscillations can modify both the energy spectrum and the flavor composition of propagated neutrinos.
If the mass-squared differences $\delta m^2_k$ are different for each mass state pair $k$, the different flavor states $\alpha$ will be affected by different combinations of oscillations, introducing new flavor dependence to the neutrino spectrum. 
Alternatively, if the mass-squared differences are all the same $\delta m^2_k = \delta m^2$, there will be no flavor-dependent effects, but the magnitude of the oscillation effects will be maximized for all flavors.

To be sensitive to both energy-only and flavor-dependent effects we combine IceCube results from two statistically independent datasets, \texttt{Cascades} and \texttt{ESTES}, that select events with different morphologies.
The \texttt{Cascades} sample~\cite{IceCube:2020acn} is a selection of cascade events collected over 6.5 years of data-taking. 
The \texttt{ESTES} sample~\cite{IceCube:2024fxo} is a selection of track events whose production vertex is within the detector volume (``starting tracks") collected over 10.3 years of data-taking. 
Since tracks in IceCube are produced primarily by muon neutrinos, with a subdominant contribution from tau neutrinos that decay to muons, whereas cascades are produced by all flavors in comparable fractions, these two selections are sensitive to different flavor combinations.

Additionally, IceCube has recently published its own \texttt{CombinedFit} analysis of the astrophysical flux, which combines an updated version of the \texttt{Cascades} sample (11.5 years) with a northern sky track selection~\cite{IceCube:2025tgp,IceCube:2025ewu}. 
Due to the increased livetime and better control of systematics, this new result is much more sensitive to the shape of the astrophysical flux.
However, because the result is reported jointly for the cascade and track event selections, it cannot be used to study flavor-dependent effects. 
We therefore study only the equi-splitting $\delta m^2_k = \delta m^2$ scenario with the \texttt{CombinedFit} results.

IceCube's analyses model the astrophysical neutrino spectrum
by splitting the spectrum piecewise into a series of energy bins, each described by an $E^{-2}$ power-law with independent normalization.
The analyses then report each segment normalization and their $\pm 1\sigma$ variations. 
These piecewise fluxes can be converted back into the most-likely number of astrophysical neutrinos per-bin by taking the product with the detector acceptance:
\begin{equation}
    N_i^S = T^S \int_{E_i}^{E_{i+1}} dE \left[ E^{-2} \Phi_i^S \sum_\alpha A_\alpha^S(E)\right].
\end{equation}
In this expression, the index $i$ indicates the energy bin, the index $S$ indicates the sample, $T^S$ is the total livetime of the sample, $\Phi_i^S$ is the reported per-bin normalization, and $A_\alpha^S$ is the flavor-dependent, zenith-averaged effective area of the sample.
Similarly, we can calculate approximate $\pm 1\sigma$ variations on the number of events $N_i^S$ by adjusting the per-bin normalization $\Phi_i^S$. 
These approximate observed event counts can then be compared to the predictions of an astrophysical flux model with QD oscillations:
\begin{equation}
\mu_i^S(\delta m^2_k, \vec{\eta}) = T^S \int_{E_i}^{E_{i+1}} dE\; \left[ \sum_\alpha \Phi_\alpha(E|\delta m^2_k, \vec{\eta}) A_\alpha^S(E) \right].
\end{equation}
The flux model $\Phi_\alpha(E)$ depends on the QD parameters, $\delta m^2_k$, as well as on the assumed source redshift distribution $\rho(z)$, the source flavor ratio $f_\alpha$, and the spectrum at the source, $\phi$, which itself depends on additional parameters, e.g., a spectral index.
We collectively denote these as the flux parameters $\vec{\eta}$, and marginalize over them in our analysis. 
When calculating flavor-dependent oscillation probabilities using~\Cref{eq:prob}, we use the PMNS matrix values reported in Ref.~\cite{Esteban:2020cvm}.
Moreover, we use the results of Ref.~\cite{Planck:2018vyg} for the parameters of our $\Lambda$CDM cosmology.

To determine preferred and disfavored regions of $\delta m^2$ parameter space, relative to the non-QD hypothesis, we use the following test statistic:
\begin{equation}
    \textrm{TS}(\delta m^2_k) = \min_{\vec{\eta}} \sum_S \sum_i -2\log\mathcal{L}_i\left(\mu_i^S(\delta m^2_k, \vec{\eta})\right),
\end{equation}
where $\mathcal{L}_i$ is an approximation of the IceCube likelihood function.
We construct this approximation differently in two regimes: for lower energy bins with $E_i<\SI{100}\TeV$, we use a Gaussian likelihood with mean $N_i^S$ and standard deviation $\sigma_i^S$, while for higher energy bins above $\SI{100}\TeV$ we use a calibrated Poisson likelihood, which is scaled so that $-2\log\mathcal{L}_i(N_i^S) = 0$ and $-2\log\mathcal{L}_i(N_i^S \pm \sigma_i^S) = 1$.
This likelihood approximation yields good agreement with IceCube results: it accurately captures the low-energy regime, where the data are dominated by backgrounds and systematics, and remains reliable at high energies, where statistical uncertainties prevail.
More details on this likelihood function, as well as a comparison of published IceCube fits and the results using our likelihood, can be found in the Appendix~\ref{sec:llh_construction}.
Assuming Wilks' theorem, we construct our confidence regions by assuming that the difference $\Delta \textrm{TS}(\delta m^2_k) = \textrm{TS}(\delta m^2_k) - \textrm{TS}(0)$, with the appropriate sign for preference or exclusion, follows a $\chi^2$ distribution.

\section{Results: equal QD squared-mass differences.}

\begin{table*}[h!]
    \renewcommand{\arraystretch}{1.2}
    \centering
    \begin{tabularx}{\textwidth}{lXr}
        \toprule
        Sample(s) & Redshift dist. $\rho(z)$ & $3\sigma$ region(s) [$\textrm{eV}^2$] \\
        \midrule
        \texttt{CombinedFit}                        & SFRD~\cite{Elias-Chavez:2018dru} & $(5.0-7.5) \times 10^{-19}$ \\
        \texttt{Cascades} + \texttt{ESTES} $\quad$  & SFRD~\cite{Elias-Chavez:2018dru} & $(5.9-7.9) \times 10^{-19}$ \\

        \texttt{CombinedFit}         & BL~Lac~\cite{Groth:2025aan} & $(2.4-3.0), (3.5-7.2), (18-35) \times 10^{-19}$ \\
        \texttt{CombinedFit}         & FSRQ~\cite{Groth:2025aan}   & $(3.5-7.2), (19-35) \times 10^{-19}$ \\
        \texttt{CombinedFit}         & LL~AGN~\cite{Groth:2025aan} & \\
        \texttt{CombinedFit}         & RQ~AGN~\cite{Groth:2025aan} & $(4.9-7.5) \times 10^{-19}$ \\

        \bottomrule
    \end{tabularx}
    \caption{
        \textbf{\textit{Limits on the squared-mass difference.}}.
        These results assume the source emission spectrum is a broken power-law (BPL).
    }
    \label{tab:excl}
\end{table*}

\begin{figure*}[ht!]
    \centering \includegraphics[width=\textwidth]{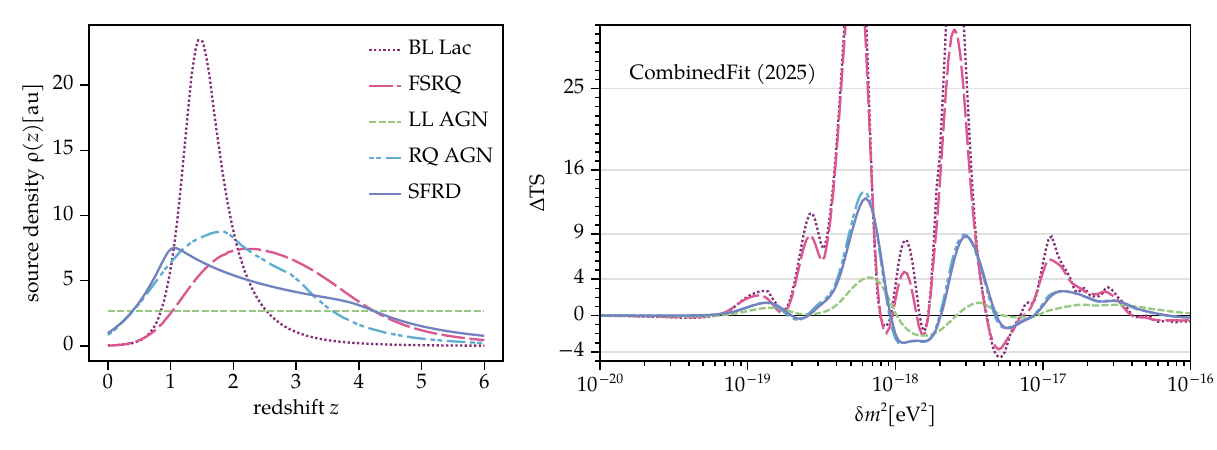}
    \caption{
        \textbf{\emph{Constraints on the QD mass-squared difference, under different choices for the source redshift distribution $\rho(z)$.}} 
        \emph{Left:} Redshift distributions of four source populations from Ref.~\cite{Groth:2025aan}, as well as the SFRD of Ref.~\cite{Elias-Chavez:2018dru}. Note that as discussed in the text, models in which neutrino sources are distributed uniformly over redshift, such as that of the LL~AGNs in this plot, are disfavored by source studies~\cite{Capel:2020txc}.
        \emph{Right:} The resulting $\textrm{TS}$ curves calculated based on the \texttt{CombinedFit} flux points, using a BPL emission spectrum.
    }
    \label{fig:zdists}
\end{figure*}

\begin{figure*}[p]
    \centering    
    \includegraphics[width=\textwidth]{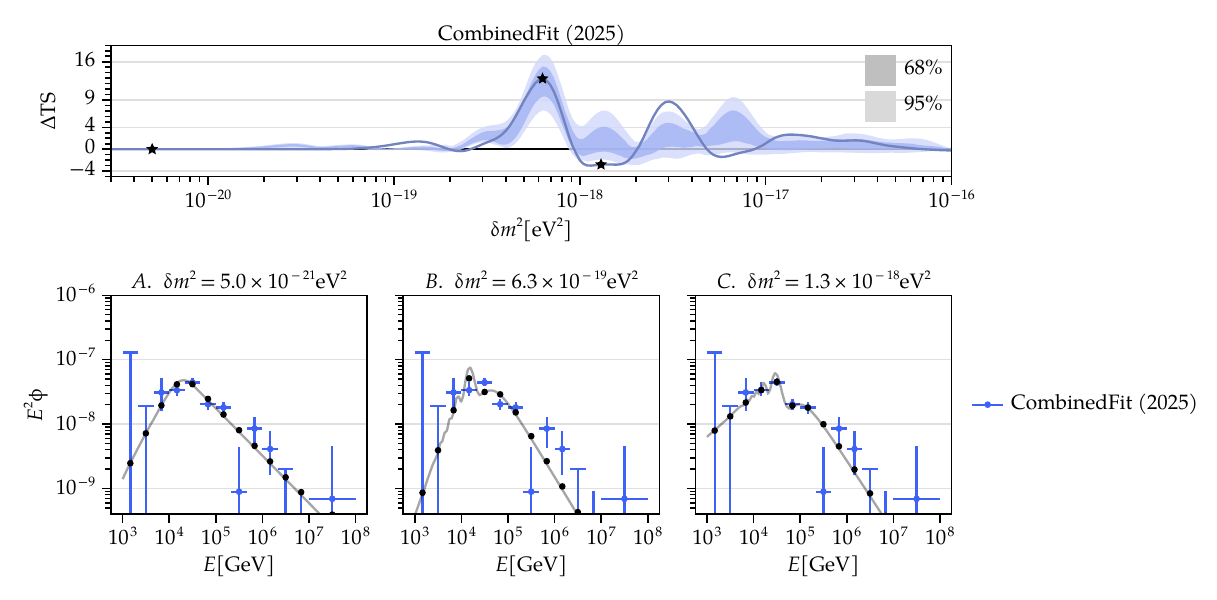}
    \caption{
        \textbf{\emph{Flux predictions and IceCube CombinedFit measurements.}}
        \emph{Below:}
        The best-fit QD flux hypotheses (grey lines) for three different values of the mass splitting $\delta m^2$, compared with the \texttt{CombinedFit} piecewise flux. The black points indicate the predicted piecewise flux in each bin according to the QD model. 
        \emph{Above:} The test statistic difference as a function of $\delta m^2$. 
        The shaded bands indicate the 68\% and 95\% most probable regions, generated by sampling from the published \texttt{CombinedFit} best-fit BPL flux.
    }
    \label{fig:fits_CF}
\end{figure*}
%

\begin{figure*}[p]
    \centering    
    \includegraphics[width=\textwidth]{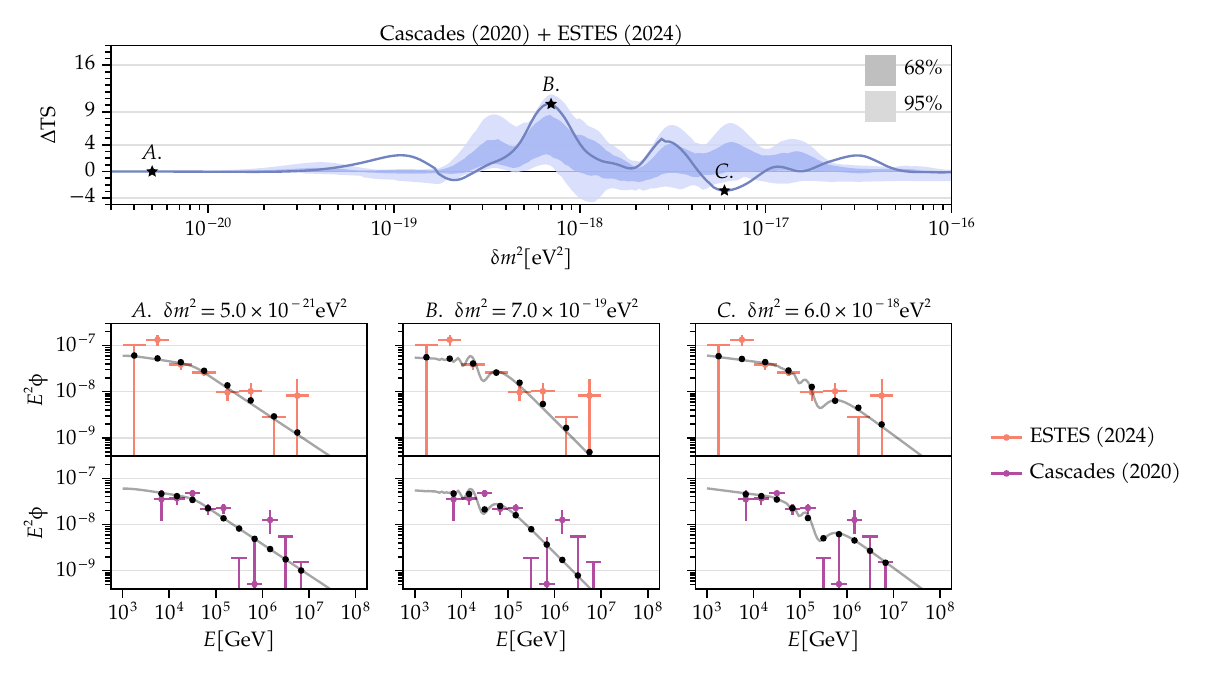}
    \caption{
        \textbf{\emph{Flux predictions and IceCube Cascades + ESTES measurements.}}
        The best-fit QD flux hypotheses (grey lines) for three different values of the mass splitting $\delta m^2$, compared with the \texttt{Cascades} and \texttt{ESTES} piecewise fluxes.
        \emph{Above:} The test statistic difference as a function of $\delta m^2$. 
        The shaded bands indicate the 68\% and 95\% most probable regions, generated by sampling from the published \texttt{CombinedFit} best-fit BPL flux.
    }
    \label{fig:fits_C20pE}
\end{figure*}
%

We begin with our results for the energy-only analysis, in which we assume a single mass-squared difference, $\delta m^2$, for all three generations.
In this scenario, the initial flavor ratio has no effect on the QD signal, and we therefore assume a pion-decay production mechanism.
For our main result, shown in~\Cref{fig:main}, we assume a broken power-law flux hypothesis for $\phi$, which is strongly preferred over a single power law in the most recent IceCube astrophysical flux results~\cite{IceCube:2025ewu, IceCube:2025tgp}, and redshift distribution function $\rho(z)$ given by the star formation rate density model of~\cite{Elias-Chavez:2018dru}.
In this case, we find that IceCube's \texttt{CombinedFit} results disfavor $\delta m^2 \in [5.0-7.5] \times 10^{-19}~\textrm{eV}^2$ at the $3\sigma$ level, and our combination of the \texttt{Cascades} and \texttt{ESTES} results disfavor a similar region $\delta m^2 \in [5.9-7.9] \times 10^{-19}~\textrm{eV}^2$ at the same level.
These statements reflect the fact that the piecewise fluxes reported by IceCube are well described by an unbroken power law in the energy range just below the break at 50~TeV, as can be seen from the lower middle panels in~\Cref{fig:fits_CF,fig:fits_C20pE}.
The constraints we obtain are stronger than the projected sensitivities of individual IceCube point sources~\cite{Carloni:2022cqz}, shown in~\Cref{fig:main} by the dark gray dashed curve, and are slightly stronger than those obtained from SN1987A neutrino data~\cite{Martinez-Soler:2021unz}.
We do not find any significant preference for a QD hypothesis.

To characterize the significance of our results, we generated Brazil bands by generating 100 pseudo-experiments assuming the best-fit BPL flux reported in IceCube's \texttt{CombinedFit} results~\cite{IceCube:2025tgp}. 
We find that the disfavored region we get from the published results is contained in the 95\% band.
Elsewhere, excursions from the band correspond to individual bins that deviate from the BPL expectation.
The $\Delta \textrm{TS}$ curve based on the \texttt{Cascades} and \texttt{ESTES} data shows a slight minimum around $\SI{6e-18}{\eV\squared}$ which is not contained within the 95\% band.
This is due to the deficit of events in the two bins in the \texttt{Cascades} results just below 1~PeV. 
A similar but less significant effect can also be seen in the \texttt{CombinedFit} curve.
The oscillatory patterns in the $\Delta \textrm{TS}$ curve are artifacts of the binned representation of the astrophysical fluxes and can be improved in internal experimental analyses.

When we change the choice of redshift distribution function $\rho(z)$, we find that the strength at which we can disfavor portions of QD parameter space can change, but the regions where significant conclusions are drawn remain substantially the same, as can be seen in~\Cref{fig:zdists}.
For this study, we considered four physical redshift evolution models from~\cite{Groth:2025aan}, representing four classes of neutrino sources: BL~Lacs, Flat~Spectrum Radio Quasars (FSRQs), Low Luminosity (LL)~AGNs, and Radio-Quiet (RQ) AGNs. 
These models also represent a range of possible redshift scalings: the BL~Lac and FSRQ models both scale as $(1+z)^5$ for small $z$, whereas the RQ~AGN model scales, like the SFRD, as $(1+z)^3$, and the LL~AGN model is flat in $z$.
This is important because, as can be seen in~\Cref{fig:oscs}, the scaling with $z$ determines the source distribution over the QD oscillation length, $L_\textrm{eff}$, and thus the extent to which the QD disappearance signal is resolved or smeared out.

We find that the models with similar scaling behavior result in very similar TS curves, and thus similar excluded regions (see~\Cref{tab:excl}). 
The BL~Lac and FSRQ models, which grow as $(1+z)^5$, both very strongly exclude regions around $(3.5-7.2)\times 10^{-19}$ and $(1.8-3.5)\times 10^{-18}~\textrm{eV}^2$, due to non-observation of  disappearance features at around 50~TeV or 100~TeV.
Conversely, the RQ~AGN and SFRD models, which both grow as $(1+z)^3$, predict less pronounced disappearance features and only disfavor $\delta m^2 \in (5-7.5) \times 10^{-19}~\textrm{eV}^2$ at the $3\sigma$ level.
Finally, the LL~AGN model, which is flat in $z$, produces a very limited disappearance signal, and thus cannot disfavor any $\delta m^2$ region at $3\sigma$.
However, we note that such flat distributions appear to be disfavored by external analyses of IceCube's high-energy neutrino data, since they would predict more bright sources than have been observed~\cite{Capel:2020txc}.
Thus, when restricting ourselves to source distributions preferred by neutrino source searches, the astrophysical spectrum remains sensitive to long-baseline oscillations.

\begin{figure}[ht!]
    \centering \includegraphics[width=\linewidth]{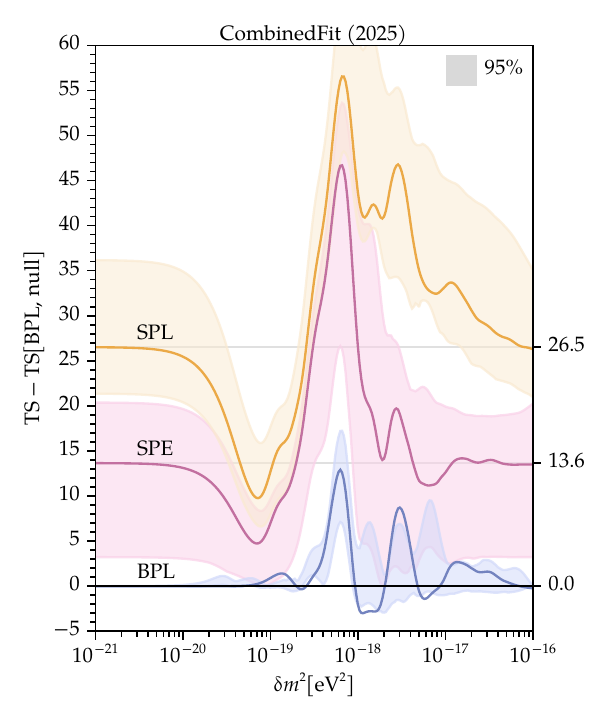}
    \caption{
        \textbf{\emph{The test statistic difference between QD models with different source emission models $\phi$ and the BPL null.}} 
        The curves shown are calculated based on the \texttt{CombinedFit} data; the bands indicate the regions contained by 95\% of realizations from the published \texttt{CombinedFit} best-fit BPL flux.
    }
    \label{fig:fmods_CF}
\end{figure}

We also considered how our results are affected by different choices for the source emission spectrum $\phi$. 
For this comparison, we consider three analytic functions commonly used to characterize the astrophysical flux: a single power-law (SPL), a single power-law with an exponential cutoff at high-energies (SPE), and a broken power-law (BPL).
The three TS curves, as well as 95\% Brazil bands generated under the \texttt{CombinedFit} BPL flux, are shown in~\Cref{fig:fmods_CF}.
Our fit reproduces the $>4\sigma$ preference for BPL over an SPL reported by IceCube's \texttt{CombinedFit} analysis~\cite{IceCube:2025tgp} in the null limit ($\delta m^2 \to\ 0$). 
We find that this gap between the spectral hypotheses cannot be fully alleviated by introducing QD oscillations.
The best SPL+QD model fit produces a $\textrm{TS} = 18.93$ for 10 degrees of freedom, which is a considerably worse fit than the BPL result, $\textrm{TS} = 9.17$ for 9 degrees of freedom (see~\Cref{tab:bestfit}).
We also observe that the tension with the SPL hypothesis substantially worsens the QD+SPL fit in the parts of parameter space that are disfavored, above $\simeq \SI{3e-19}{\eV\squared}$.
The TS curves for all three spectral shapes are consistent with the 95\% Brazil band produced under the \texttt{CombinedFit} BPL best-fit spectrum.

\begin{figure}[ht!]
    \centering \includegraphics[width=\linewidth]{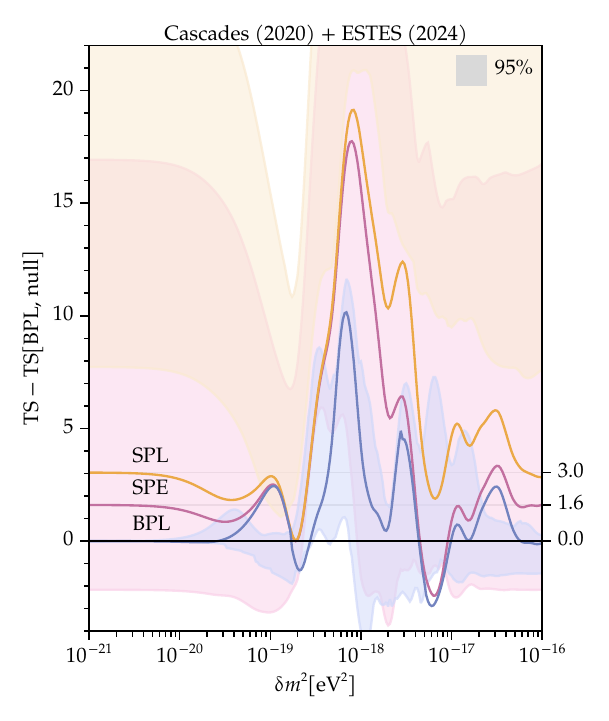}
    \caption{
        \textbf{\emph{The test statistic difference between QD models with different source emission models $\phi$ and the BPL null.}} 
        The curves shown are calculated based on the combined \texttt{Cascades} and \texttt{ESTES} data; the bands indicate the regions contained by 95\% of realizations from the published \texttt{CombinedFit} best-fit BPL flux.
    }
    \label{fig:fmods_C20pE}
\end{figure}

The BPL fit to the combined \texttt{Cascades} and \texttt{ESTES} flux measurements, shown in~\Cref{fig:fmods_C20pE}, improves over a SPL by only three TS units. 
In this case, we find that the SPL+QD and SPE+QD models, at $\delta m^2 = \SI{1.9e-19}{\eV\squared}$ are able to describe the reported fluxes equally as well as a BPL without QD effects.
Additionally, a mass-squared difference of $\delta m^2 \simeq \SI{6e-18}{\eV\squared}$ is able to improve the fit in the energy bins near 1~PeV, where there is a deficit in~\texttt{Cascades}.
However, neither of these improvements are statistically significant.
As we found in the~\texttt{CombinedFit} study, the additional degrees of freedom in the BPL hypothesis can absorb some of the tension induced by QD mass squared differences between $3 \times 10^{-19}$ and $\SI{5e-18}{\eV\squared}$.

We conclude that the choice of emission spectrum $\phi$ does not change the region in $\delta m^2$ this analysis is sensitive to, but can modify the strength of statements in either direction.
For this reason, we chose the BPL model, which in the null scenario describes the data best, and produces the most conservative limits, for our main result.


\begin{table*}[h!]
    \centering
    \begin{tabularx}{\textwidth}{lXXXcc}
        \toprule
        Sample(s) & Flux Model $\phi$ & $\delta m^2 [\textrm{eV}^2]$ & TS & dof$\quad$ & p-value \\
        \midrule

        \texttt{CombinedFit} &&&& \\
        & SPL & \emph{null}             & 35.67 & 11 & 0.02\%  \\
        & SPL & $7.2 \times 10^{-20}$   & 18.93 & 10 & 4.11\%  \\
        & SPE & \emph{null}             & 22.82 & 10 & 1.14\%  \\
        & SPE & $6.9 \times 10^{-20}$   & 13.91 & 9  & 12.57\%  \\
        & BPL & \emph{null}             & 9.17  & 9  & 42.20\% \\
        & BPL & $1.2 \times 10^{-18}$   & 6.17  & 8  & 62.84\% \\

        \texttt{Cascades} + \texttt{ESTES} $\quad$ &&&& \\
        & SPL & \emph{null}             & 28.10 & 16 & 3.07\% \\
        & SPL & $1.9 \times 10^{-19}$   & 25.10 & 15 & 4.87\% \\
        & SPE & \emph{null}             & 26.67 & 15 & 3.16\% \\
        & SPE & $6.3 \times 10^{-18}$   & 22.64 & 14 & 6.63\% \\
        & BPL & \emph{null}             & 25.06 & 14 & 3.39\% \\
        & BPL & $6.0 \times 10^{-18}$   & 22.18 & 13 & 5.26\% \\
        \bottomrule
    \end{tabularx}
    \caption{\textbf{\textit{Best fit parameters, for each combination of IceCube results used, assuming a single mass-squared difference $\delta m^2$.}} These results use the SFRD~\cite{Elias-Chavez:2018dru} for the source redshift distribution $\rho(z)$.}
    \label{tab:bestfit}
\end{table*}

\section{Results: multiple QD squared-mass differences.}
We search for flavor-dependent QD signals by combining the \texttt{Cascades} and \texttt{ESTES} measurements.
To restrict the available parameter space, we consider only scenarios with two distinct squared mass differences, e.g. $\delta m^2_1 = \delta m^2_2 \neq \delta m^2_3$.
For this study, we assume a BPL for the emission spectrum, and use the SFRD for the source distribution over redshift.
We consider three possible scenarios for the flavor ratio at the source: pion decay (1,2,0), muon-damping (0,1,0), and neutron-dominated (1,0,0).
In both the pion decay and neutron-dominated scenarios, oscillations induced by $\delta m^2_1$ disproportionately affect cascades, whereas those induced by $\delta m^2_3$ disproportionately affect tracks. 
In the muon-damped scenario, the electron fraction in both tracks and cascades is substantially reduced, so $\delta m^2_1$ oscillations have very little impact on either morphology.

\begin{figure*}[h!]
    \centering \includegraphics[width=\textwidth]{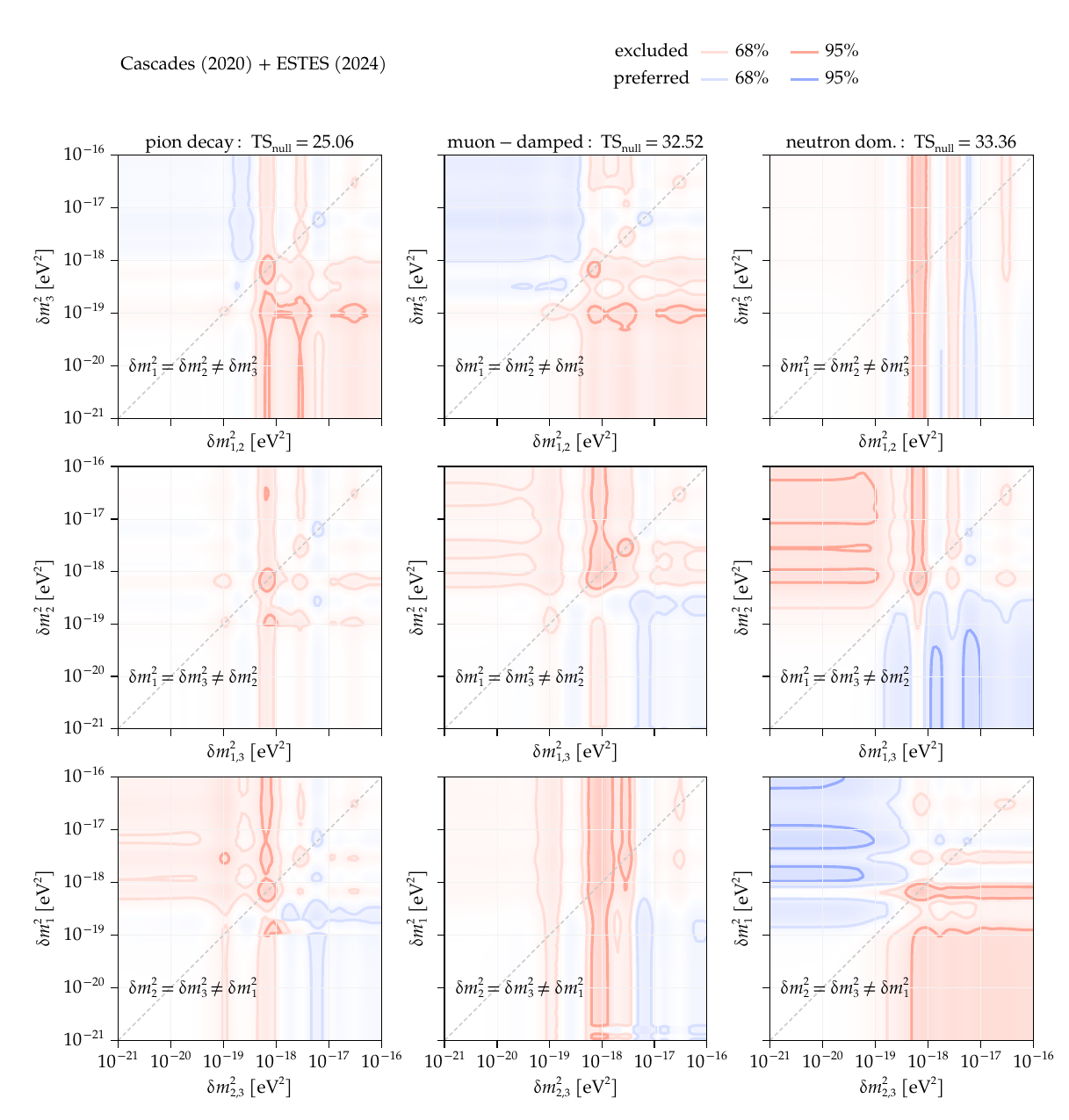}
    \caption{
        \textbf{\emph{Disfavored and preferred regions in 2-dimensional QD parameter space.}} Left to right, the three columns correspond to the initial flavor ratios of the pion decay, muon damped, and neutron dominated scenarios.
        These results use a source distribution over redshift that follows the SFRD, and assume the a BPL emission spectrum.
    }
    \label{fig:grid2D}
\end{figure*}

\begin{figure*}[h!]
    \centering \includegraphics[width=\textwidth]{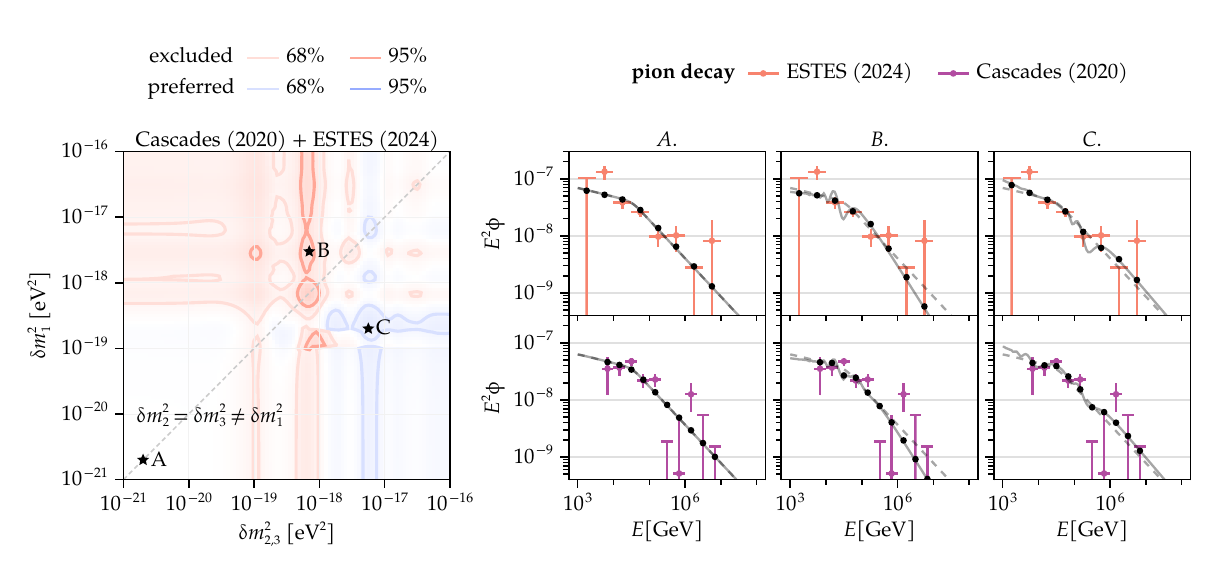}
    \caption{
        \textbf{\emph{Two-parameter QD models in the pion decay production scenario.}}
        \emph{Left:} The test statistic difference as a function of $\delta m^2_1$ and $\delta m^2_2 = \delta m^3$, assuming an initial flavor ratio produced by pion decays (1,2,0).
        \emph{Right:} The best-fit QD flux hypotheses (solid grey curves) for the three points in two-dimensional QD parameter space indicated, compared with the \texttt{Cascades} and \texttt{ESTES} piecewise fluxes. The black points indicate the predicted piecewise flux in each bin according to the QD model. The dashed grey curves indicate the null best fit.
    }
    \label{fig:fits2D_pi}
\end{figure*}

\begin{figure*}[h!]
    \centering \includegraphics[width=\textwidth]{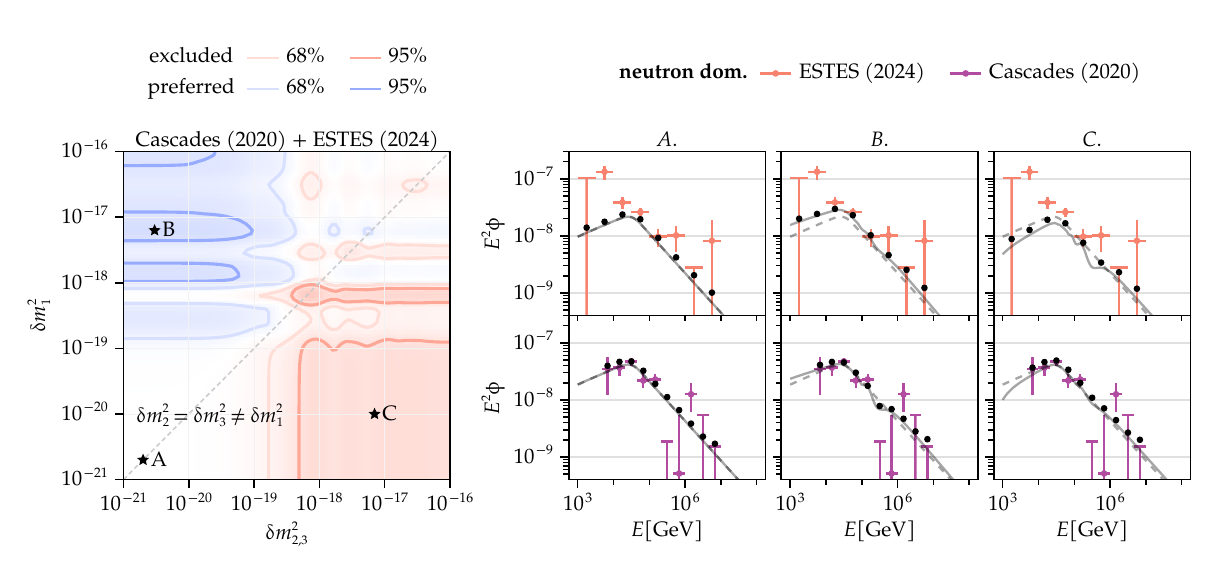}
    \caption{
        \textbf{\emph{Two-parameter QD models in the neutron-dominated production scenario.}}
        \emph{Left:} The test statistic difference as a function of $\delta m^2_1$ and $\delta m^2_2 = \delta m^3$, assuming an initial flavor ratio dominated by neutron decays (1,0,0).
        \emph{Right:} The best-fit QD flux hypotheses for the three points in two-dimensional QD parameter space indicated, compared with the \texttt{Cascades} and \texttt{ESTES} piecewise fluxes. 
    }
    \label{fig:fits2D_n}
\end{figure*}

The full set of results, for the three different source flavor assumptions and three different choices for the distinct $\delta m^2$, are shown in~\Cref{fig:grid2D}, and the best-fit parameters are listed in~\Cref{tab:bestfit2D}.
With the additional degree of freedom, we do not find any significant preferences or exclusions at the $3\sigma$ level for these scenarios.
In the absence of QD effects, the muon-damped and neutron-dominated scenarios, which do not predict equal flavor ratios at Earth, produce significantly worse fits than the pion-decay scenario. 
We find that in the neutron-dominated case, a quasi-Dirac model with $\delta m^2_1 = \SI{6.3e-18}{\eV\squared}, \delta m^2_2 = \delta m^2_3 = \SI{1e-21}{\eV\squared}$ can improve the fit at the $2\sigma$ level.
As shown in~\Cref{fig:fits2D_n}, the oscillations induced by $\delta m^2_1$ produce a slight dip in the effective \texttt{Cascades} flux around 500~TeV, without modifying the muon neutrino dominated \texttt{ESTES} flux.
However, the TS of the resulting model is only comparable to that of the null pion-decay scenario, at the cost of two additional degrees of freedom.
A similar combination of parameters, shown in~\Cref{fig:fits2D_pi}, is able to improve the fit in a pion decay production scenario, although only at the $1.4\sigma$ level.

\begin{table*}[h!]
    \centering
    \begin{tabularx}{\textwidth}{llcXXXccc}
        \toprule
        Flavor Ratio && Flux Model $\phi$ & $\delta m^2_1 [\textrm{eV}^2]$ & $\delta m^2_2 [\textrm{eV}^2]$ & $\delta m^2_3 [\textrm{eV}^2]$ &  TS & dof$\quad$ & p-value \\
        \midrule
        Pion decay   & (1,2,0) & BPL & \emph{null} & \emph{null} & \emph{null} & 25.06 & 14 & 3.39\% \\
        Muon-damped  & (0,1,0) & BPL & \emph{null} & \emph{null} & \emph{null} & 32.52 & 14 & 0.34\% \\
        Neutron dom. & (1,0,0) & BPL & \emph{null} & \emph{null} & \emph{null} & 33.35 & 14 & 0.26\% \\
        Pion decay   & (1,2,0) & BPL & $2.0 \times 10^{-19}$  & $5.6 \times 10^{-18}$ & $=\delta m^2_2$          & 20.00 & 12 & 6.71\% \\
        Muon-damped  & (0,1,0) & BPL & $2.5 \times 10^{-20}$  & $=\delta m^2_1$       & $5.6 \times 10^{-18}$    & 27.12 & 12 & 0.74\% \\
        Neutron dom. & (1,0,0) & BPL & $6.3 \times 10^{-18}$  & $1.0 \times 10^{-21}$ & $=\delta m^2_2$          & 25.63 & 12 & 1.21\% \\
        \bottomrule
    \end{tabularx}
    \caption{
        \textbf{\textit{Best fit parameters for fits assuming two distinct mass-squared differences.}} 
        These results are based on the combined \texttt{Cascades} and \texttt{ESTES} flux points, and use the SFRD~\cite{Elias-Chavez:2018dru} for the source redshift distribution $\rho(z)$.
    }
    \label{tab:bestfit2D}
\end{table*}

\section{Conclusions.}
Neutrino telescopes provide a unique opportunity to uncover the origin of neutrino masses.
In particular, the quasi-Dirac (QD) neutrino model, which is motivated by string theory scenarios, produces observable signatures in the spectra of astrophysical neutrinos.
In this work, we have shown that QD oscillations remain observable from an extended populations of sources.
These produce observable disappearance signatures in the spectra and can impact the flavor ratio of astrophysical neutrinos.

Assuming that neutrinos are produced by a population of sources distributed over redshift $z$ according to the star formation rate density (SFRD), and that all sources emit neutrinos according to a broken power-law spectrum, we find that the most recent IceCube measurement of the astrophysical spectrum disfavors, at the $3\sigma$ level, a single QD mass-squared difference $\delta m^2$ between $(5 \times 10^{-19}, 8 \times 10^{-19})~\textrm{eV}^2$. 
Our sensitivity to this region is increased when we consider alternate possibilities for the source evolution in $z$, or alternate shapes for the emitted source spectrum.  

When we consider a model with two distinct mass splittings, which would produce energy-dependent modifications to the astrophysical flavor ratio, we find that QD models can alleviate the tension between track and cascade based measurements of the astrophysical flux, producing a $1\sigma$ improvement in a pion decay production scenario and a $2\sigma$ improvement in a neutron-dominated scenario.

Future observations of the astrophysical flux, leveraging more years of data and better reduced  systematic uncertainties, will further improve our understanding of its shape and flavor dependence, and thus further extend the sensitivity to neutrino mass models.
Additionally, point source analyses are continuing to determine the origins of high-energy astrophysical neutrinos, constraining the properties of the source population and improving our sensitivity to propagation effects such as QD-driven oscillations.
These two beams are complementary.
Hints of ultra-long baseline observations can be first observed by features in the diffuse, all-sky neutrion emission.
Such signals could then be confirmed by the study of the neutrino yield from multiple sources.

\section{Acknowledgments.}
We thank Subir Sarkar for useful comments on the draft. We also thank Peter Denton and Andre de Gouvea for useful discussion.  
KC is supported by the NSF Graduate Research Fellowship under Grant No. 2140743, and the Research Corporation for Science Advancement Cottrell Scholar award. 
The work of YP was supported by the S\~{a}o Paulo Research Foundation (FAPESP) Grant No. 2023/10734-3 and 2023/01467-1, and by the National Council for Scientific and Technological Development (CNPq) Grant No. 151168/2023-7.
CAA are supported by the Faculty of Arts and Sciences of Harvard University, the National Science Foundation, the Research Corporation for Science Advancement, and the David \& Lucile Packard Foundation. 
The work of BD is partly supported by the U.S. Department of Energy under grant No. DE-SC0017987. 
SJ would like to acknowledge support from the Department of Atomic Energy, Government of India, for the Harish-Chandra Research Institute.

\bibliography{main.bib}

\clearpage
\pagebreak
\appendix
\onecolumngrid

\ifx \standalonesupplemental\undefined
    \setcounter{page}{1}
    \setcounter{figure}{0}
    \numberwithin{equation}{section}
\fi

\renewcommand{\thepage}{Supplemental Methods and Tables --- S\arabic{page}}
\renewcommand{\figurename}{SUPPL. FIG.}
\renewcommand{\tablename}{SUPPL. TABLE}
\renewcommand{\theequation}{A\arabic{equation}}
\clearpage

\begin{center}
\textbf{\large Supplemental Material}
\end{center}

\section{Likelihood construction and compatibility with IceCube results}
\label{sec:llh_construction}
As noted in the main text, we approximate the IceCube likelihood function by the following function
\begin{equation}
    \textrm{TS}(\delta m^2_k) = \min_{\vec{\eta}} \sum_S \sum_i -2\log\mathcal{L}_i\left(\mu_i^S(\delta m^2_k, \vec{\eta})\right).
\end{equation}
This function needs to have the right statistical coverage as reported by the IceCube results, while also have the proper limiting cases, which are distinct in signal and background dominated regimes.
At sub-100~TeV energies, as the flux normalization goes to zero the likelihood should converge to a constant value, since this regime is background dominated.
At supra-100~TeV energies, the data is dominated by astrophysical neutrinos, with negligible background contribution, and thus the flux normalization is proportional to the rate.
This implies that the in this regime the likelihood should behave asymptotically like a Poisson likelihood.
Explicitly, for $E_\nu < \SI{100}\TeV$,
\begin{equation}
    -2\log\mathcal{L}_i\left(\mu_i^S(\delta m^2_k, \vec{\eta})\right) = \frac{\left(N_i^S - \mu_i^S(\delta m^2_k, \vec{\eta})\right)^2}{(\sigma_i^{\pm,S})^2},
\end{equation}
i.e., a Pearson-$\chi^2$, and where $\sigma_i^{\pm,S}$ corresponds to the upper error $(\sigma_i^{+,S})$ on the data when $\mu_i^S > N_i^S$ and the lower errors $(\sigma_i^{+,S})$ otherwise.
For $E_\nu > \SI{100}\TeV$,
\begin{equation}
    -2\log\mathcal{L}_i\left(N_i^S|\mu_i^S(\delta m^2_k, \vec{\eta})\right) = -2 \left[\frac{\log\mathcal{L}^{\rm Poisson}\left(N_i^S |\mu_i^S(\delta m^2_k, \vec{\eta})\right) - \log\mathcal{L}^{\rm Poisson}\left(N_i^S | N_i^S\right)}{
    \log\mathcal{L}^{\rm Poisson}\left(N_i^S | N_i^S\pm\sigma_i^{\pm,S}\right) - \log\mathcal{L}^{\rm Poisson}\left(N_i^S|N_i^S\right)
    }
    \right],
\end{equation}
where $\mathcal{L}^{\rm Poisson}\left(N|\mu\right)$ is the probability of observing $N$ given expected $\mu$ and the ``$+$'' and $\sigma_i^{+,S}$ is used for $\mu_i^S >N_i^S$ while the corresponding minus sign pair in the contrary case.

As can be seen in~\Cref{tab:icecube_validation}, using this likelihood approximation, we reproduce all relevant results published by the experiment within uncertainties.

\begin{table}[h]
    \renewcommand{\arraystretch}{1.2}
    \centering
    \begin{tabularx}{\linewidth}{XXlllll}
        \toprule
        Fit & Sample & Flux Model &&& \\
        \midrule
        &&                            SPL & $\gamma$                 & $\phi_0$                 && \\
        \cmidrule(lr){3-5}
        IceCube     & \texttt{Cascades}    && $2.53^{+0.07}_{-0.07}$   & $1.66^{+0.25}_{-0.27}$   && \\
        This work   & \texttt{Cascades}    && $2.41$                   & $1.95$                   && \\
        IceCube     & \texttt{ESTES}       && $2.58^{+0.1}_{-0.09}$    & $1.68^{+0.19}_{-0.22}$   && \\
        This work   & \texttt{ESTES}       &&  $2.57$                  & $1.60$                   && \\
        IceCube     & \texttt{CombinedFit} && $2.52^{+0.036}_{-0.038}$ & $1.8^{+0.13}_{-0.16}$    && \\
        This work   & \texttt{CombinedFit} && $2.35$                   & $1.83$                   && \\
        &&&&&&\\
        &&                            SPE & $\gamma$                & $\phi_0$ & $\log_{10}E_\textrm{cut}$& \\
        \cmidrule(lr){3-6}
        IceCube     & \texttt{Cascades}    && $2.45^{+0.09}_{-0.11}$   & $1.83^{+0.37}_{-0.31}$   & $6.4^{+0.9}_{-0.4}$       & \\
        This work   & \texttt{Cascades}    && $2.30$                   & $2.41$                   & $6.29$                    & \\
        IceCube     & \texttt{CombinedFit} && $2.386^{+0.081}_{-0.09}$ & $2.2^{+0.3}_{-0.25}$     & $6.15^{+0.37}_{-0.24}$    & \\
        This work   & \texttt{CombinedFit} && $2.19$                   & $2.40$                   & $5.99$                    & \\
        &&&&&&\\
        &&                            BPL & $\phi_0$                & $\log_{10}E_\textrm{break}$ & $\gamma_1$          & $\gamma_2$ \\
        \cmidrule(lr){3-7}
        IceCube     & \texttt{Cascades}  && $1.71^{+0.65}_{-0.29}$  & $4.6^{+0.5}_{-0.2}$   & $2.11^{+0.29}_{-0.67}$    & $2.75^{+0.29}_{-0.14}$    \\
        This work   & \texttt{Cascades}  && $2.15$                  & $4.42$                & $1.52$                    & $2.67$                    \\
        IceCube     & \texttt{ESTES}     && $1.70^{+0.19}_{-0.22}$  & $4.36$                & $2.79^{+0.3}_{-0.5}$      & $2.52^{+0.1}_{-0.09}$     \\
        This work   & \texttt{ESTES}     && $1.51$                  & $4.67$                & $2.46$                    & $2.72$                    \\
        IceCube     & \texttt{CombinedFit} && $1.77^{+0.19}_{-0.18}$  & $4.39^{+0.1}_{-0.1}$  & $1.31^{+0.51}_{-1.3}$     & $2.735^{+0.067}_{-0.075}$ \\
        This work   & \texttt{CombinedFit} && $1.75$                  & $4.36$                & $1.25$                    & $2.74$                    \\
        \bottomrule
    \end{tabularx}
    \caption{\textbf{\textit{Compatibility with IceCube results}} Best-fit parameters obtain in our analysis compared with IceCube published values and their uncertainties.}
    \label{tab:icecube_validation}
\end{table}

\end{document}